\documentclass[fleqn,usenatbib]{mnras}
\usepackage[T1]{fontenc}
\usepackage{ae,aecompl}
\usepackage{graphicx}
\usepackage{amsmath}
\usepackage{amssymb}
\usepackage{cleveref}
\usepackage{txfonts}

\newcommand{\ltsima}{$\; \buildrel < \over \sim \;$}
\newcommand{\lsim}{\lower.5ex\hbox{\ltsima}}
\newcommand{\gtsima}{$\; \buildrel > \over \sim \;$}
\newcommand{\gsim}{\lower.5ex\hbox{\gtsima}}
\newcommand{\bra}{\langle}
\newcommand{\ket}{\rangle}

\newcommand{\dd}{\mathrm{d}}

\title[Horndeski gravity with intrinsic alignments]
{Testing modified (Horndeski) gravity by combining intrinsic galaxy alignments with cosmic shear}
\author[]
{Robert Reischke$^{1}$\thanks{email: \href{mailto:reischke@astro.ruhr-uni-bochum.de}{reischke@astro.ruhr-uni-bochum.de}}, Victor Bosca$^{2,3}$, Tim Tugendhat$^2$,  Bj{\"o}rn Malte Sch{\"a}fer$^2$\thanks{e-mail: bjoern.malte.schaefer@uni-heidelberg.de}\\
$^1$ Ruhr University Bochum, Faculty of Physics and Astronomy, Astronomical Institute (AIRUB),\\ \hspace{0.15cm} German Centre for Cosmological Lensing, 44780 Bochum, Germany \\
$^2$Astronomisches Rechen-Institut, Zentrum f{\"u}r Astronomie der Universit{\"a}t Heidelberg, Philosophenweg 12, 69120 Heidelberg, Germany\\
$^3$Instituto de F\'{i}sica Te\'{o}rica UAM-CSIC, Universidad Auton\'{o}ma de Madrid,  Cantoblanco, 28049 Madrid, Spain
}

\begin{document}
\pagerange{\pageref{firstpage}--\pageref{lastpage}}
\pubyear{2021}
    \maketitle
\label{firstpage}

\begin{abstract}
We study the impact of modified gravity of the Horndeski class, on intrinsic shape correlations in cosmic shear surveys. As intrinsic shape correlations (IAs) are caused by tidal gravitational fields acting on galaxies as a collection of massive non-relativistic test particles, they are only sensitive to the gravitational potential, which forms in conjunction with the curvature perturbation. In contrast, the cosmic shear signal probes the sum of these two, i.e. both Bardeen-potentials. Combining these probes therefore constitutes a test of gravity, derived from a single measurement.

Focusing on linear scales and alignments of elliptical galaxies, we study the impact on inference of the braiding $\hat{\alpha}_B$ and the time evolution of the Planck mass $\hat{\alpha}_M$ by treating IAs as a genuine signal contributing to the overall ellipticity correlation. We find that for \textsc{Euclid}, IAs can help to improve constraints on modified gravity of the Horndeski-class by 10 per cent if the alignment parameter needed for the linear alignment model is provided by simulations. If, however, the IA needs to be self calibrated, all of the sensitivity is put into the inference of the alignment strength $D$ since there is a very strong correlation with the evolution of the Planck mass. Thus diminishing the benefit of IA for probing modified gravitational theories.  While the present paper shows results mainly for modified gravity parameters, similar deductions can be drawn for the investigation of anisotropic stresses, parameterised modifications to the Poisson-equation, the phenomenology of gravitational slip and to breaking degeneracies in a standard cosmology.
\end{abstract}


\section{Introduction}
\label{sec:intro}
The combination of different cosmological probes such as type Ia Supernovae \citep[SNIa, e.g.][]{perlmutter_discovery_1998, riess_observational_1998, perlmutter_measurements_1999, riess_type_2004, riess_new_2007, riess_large_2019}, the angular power spectra of the cosmic microwave background (CMB) anisotropies \citep[e.g.][]{hinshaw_nine-year_2013,planck_collaboration_planck_2020} and of galaxy clustering \citep[e.g.][]{cole_2df_2005,beutler_clustering_2017,satpathy_clustering_2017} led to the conclusion that the Universe is expanding in an accelerated fashion. Explaining these results within the standard cosmological model $\Lambda$CDM based on general relativity (GR) as the theory of gravity, requires a non-zero but small cosmological constant $\Lambda$ and stipulates that the bulk of gravitating matter is cold and dark.

Usually, observations are either sensitive to the cosmological background model or the perturbations on top of the former. While the background model has been explored in great detail via SNIa or Baryon Acoustic Oscillations (BAOs) the same level of detail has been achieved for the large-scale structure (LSS) very early in the Universe via observations of the CMB, painting a very consistent picture of cosmic evolution. Upcoming surveys of the LSS will provide exquisite data of the perturbed Universe and test models of gravity close to the fundamental limits of inference.

So far, general relativity has been tested on non-cosmological scales and in the weak field limit only \citep[see][for reviews,]{heavens_model_2007, jain_observational_2008, bertschinger_distinguishing_2008, berti_testing_2015}, with inconclusive answers on tensions of the data with $\Lambda$CDM \citep{giannantonio_new_2010, dossett_constraints_2015}. Observations of neutron star mergers \citep{abbott_multi-messenger_2017} are found to be consistent with GR. Constraining the sound speed of the tensorial modes to be equal to the speed of light, having profound implication on the parameter space of a variety of modified gravity models \citep[e.g.][]{baker_strong_2017,creminelli_dark_2017,ezquiaga_dark_2017,sakstein_implications_2017,lombriser_challenges_2017}. There is, however, still a lot of room and general relativity needs to be tested on cosmological scales 
{\citep[see e.g.][]{lue_differentiating_2004, laszlo_non-linear_2007, kunz_dark_2007, koyama_cosmological_2016, joyce_dark_2016, white_marked_2016}}. Quite generally, modifications to general relativity lead to very different phenomena \citep[see][for a review]{clifton_modified_2012} and influence the background expansion as well as the growth of structures \citep{zhao_searching_2009, kobayashi_evolution_2010}. Models with modifications to gravity are naturally degenerate \citep{bhattacharya_bispectrum_2012, battye_parametrizing_2013} with (clustering) dark energy models \citep[for a review of these models we refer to][]{copeland_dynamics_2006}. 

In the next decade we expect a huge step forward in LSS surveys \citep{albrecht_report_2006}. These are in particular the \textsc{Euclid} mission \citep{laureijs_euclid_2011} or \textsc{LSST} \citep{lsst_dark_energy_science_collaboration_large_2012}. Of particular interest is the weak gravitational lensing signal of the LSS, called cosmic shear \citep[see e.g.][]{bartelmann_weak_2001,hoekstra_weak_2008,kilbinger_cosmology_2015}. The latter encodes information about both structure growth, background dynamics and most importantly for modified gravity it measures the Weyl potential, i.e. both the time and space components of the metric perturbations due to the null condition of the geodesic equation. 

One of the major systematic effects in cosmic shear measurements are intrinsic alignments (IA), which mimic correlation in the shapes of neighbouring galaxies \citep[see e.g.][]{schaefer_review:_2009,joachimi_intrinsic_2010,joachimi_intrinsic_2013-1,kirk_galaxy_2015,kiessling_galaxy_2015-1,troxel_intrinsic_2015}. The exact mechanisms for all scales involved are not yet clarified and may differ for different galaxy types, yet there exist physically well-motivated models. These include tidal alignment models \citep{hirata_intrinsic_2004,hirata_intrinsic_2010,blazek_testing_2011,joachimi_intrinsic_2013, blazek_tidal_2015,tugendhat_angular_2017}, extendible to nonlinear scales \citep{blazek_beyond_2017} and models based on the halo distribution of matter \citep{vlah_eft_2019,fortuna_halo_2020}. Tidal alignment models of velocity-dispersion supported elliptical galaxies currently has the strongest observational support by a number of works \citep[e.g][]{mandelbaum_detection_2006, hirata_intrinsic_2007, joachimi_constraints_2011, okumura_gravitational_2009, johnston_kidsgama_2019}. The data on torquing of spiral galaxies, which likewise would predict ellipticity correlations, is inconclusive in simulations \citep{chisari_intrinsic_2015, tenneti_galaxy_2015, kraljic_and_2020, samuroff_advances_2020, zjupa_intrinsic_2020} and depends on the implementation of feedback and hydrodynamics which have a smaller effect on the tidal alignment due its larger correlation length. Especially \citet{zjupa_intrinsic_2020} find no quadratic response of the ellipticity to the tidal field, instead they find a linear one similar to the alignment of elliptical galaxies. 

While IAs is usually regarded as a contamination to cosmic shear measurements, they contain in principle valuable cosmological information, which can be accessed if the details and parameters of the alignment process are understood well enough. Tidal alignment of elliptical galaxies probes like gravitational lensing shear tidal gravitational fields with the subtle difference that only the gravitational potential as the metric perturbation is probed and not the sum of gravitational potential and curvature perturbation. In gravity theories where these two Bardeen potentials are not equal, gravitational slip is generated, and the motion of relativistic and non-relativistic test particles is changed. Commonly, one investigates this by combining lensing and galaxy clustering, possibly together with redshift space distortions, and gains in this way access to possible differences between the two Bardeen-potentials.

In this work, we investigate the possibility to use the intrinsic alignment signal in conjunction with the weak lensing in order to improve constraints on modified gravity theories. The advantage of this measurement is that the degeneracy of the sum of the Bardeen potentials is broken within a single measurement and that both probes are sensitive to tidal fields, derived from one or from both Bardeen-potentials, respectively. Therefore, there is no extrapolation of scales involved, which is markedly different in redshift space distortions, where the velocity field probes the first rather than the second derivatives of the gravitational potential and involves consequently perturbations with larger wave lengths.

In particular, we will focus on Horndeski gravity \citep{horndeski_second-order_1974, nicolis_galileon_2009, deffayet_$k$-essence_2011} which provides the most general second order Lagrange-density free of Ostrogradsky instabilities and naturally generates gravitational slip. We will work on large scales where a linear theory of structure formation is applicable and the tidal shearing model can be assumed to be a good description of the alignment process for elliptical galaxies. Other physical theories where our method would be applicable are cosmologies based on standard general relativity but with anisotropic stresses due to non-ideal fluids. Of course as models to be tested by a combination of lensing-induced and intrinsic ellipticity correlations one could choose purely phenomenological parameterisations extrapolating for instance the Poisson-equation by introducing $\eta$ and $\mu$ or dynamical parameters like gravitational $\varpi$ slip itself.

The structure of the paper is the following: In \cref{sect_cosmology} we summarise the basic properties of Horndeski theories of gravity and discuss the background and first order perturbation dynamics Then, in \cref{sect_slip}, we describe the sensitivity of cosmic shear and IA to gravitational slip . In \cref{sect_results} we present the results and summarise in \cref{sect_summary}.

\section{Background cosmology and linear perturbations}\label{sect_cosmology}
The most general scalar-tensor theory of gravity \citep{horndeski_second-order_1974} obeys the following Lagrange density:
\begin{equation}\label{eq:Horndeski_Lagrangian}
\mathcal{L} = \sum_{i=2}^5 L_i\left[\phi,g_{\mu\nu}\right] + L_\mathrm{m}\left[g_{\mu\nu},\psi\right],
\end{equation}
with corresponding action $S = \int\dd^4x\sqrt{-g}\mathcal{L}$, where $\dd^4x\sqrt{-g}$ is the canonical volume form, $\phi$ the additional scalar degree of freedom, $g_{\mu\nu}$ the metric and $\psi$ the matter fields. The individual terms in the Lagrange density are given by
\begin{equation}
\begin{split}
{L}_2 &= G_2 (\phi, X), \\
L_3 &= -G_3(\phi, X) \Box \phi,  \\
{L}_4 &=  G_4 (\phi, X) R + G_{4X}(\phi, X) \left[ (\Box \phi)^2 - \phi_{;\mu\nu} \phi^{;\mu\nu} \right],\\
{L}_5 &= G_5 (\phi, X) G_{\mu\nu} \phi^{;\mu\nu}  \\ 
&- \frac{1}{6}G_{5X} (\phi, X) \left[ (\Box \phi)^3 + 2 \phi_{;\mu}^{\nu} \phi_{;\nu}^{\alpha} \phi_{;\alpha}^{\mu} - 3 \phi_{;\mu\nu} \phi^{;\mu\nu} \Box \phi \right].
\end{split}
\end{equation} 
 The kinetic term of the field is labelled $X \equiv - \nabla_\nu \phi\nabla^{\nu}\phi/2$. It remains to specify the four functions $G_j$ and $K=G_2$ to characterise the theory completely. Covariant derivatives are denoted by semicolons.

Assuming linear perturbations to a Friedmann-Robertson-Walker metric, the line element can be written as
\begin{equation}
    \mathrm{d}s^2 = -\left(1+2\Phi\right)c^2\mathrm{d}t^2 + a^2(t) \left(1-2\Psi\right)\mathrm{d}\boldsymbol{x}^2\;,
\end{equation}
where the scalar perturbations $\Phi$ and $\Psi$ are called the Bardeen potentials. These gauge invariants satisfy $\Phi = \Psi$ in general relativity.
In \citet{bellini_maximal_2014} it was shown that the evolution of linear perturbations in Horndeski theories can be completely characterized by free functions depending on time only by virtue of an effective field theory approach:
\begin{equation}
\begin{split}
M_*^2 & =\  2\left(G_4 -2X G_{4X} + XG_{5\phi}-\phi HXG_{5X}\right)\, , \\
HM_*^2\alpha_\mathrm{M} & \equiv \frac{\mathrm{d}M_*^2}{\mathrm{d}t}\, , \\
H^2M^2_*\alpha_\mathrm{K} & \equiv \ 2X\left(K_X + 2XK_{XX} - 2G_{3\phi} - 2XG_{3\phi}\right)\\
& +12\dot\phi XH\left(G_{3X}+XG_{3XX} -3G_{4\phi X} - 2XG_{4\phi XX}\right) \\
& +12X H ^2\left(G_{4X} +8XG_{4XX} + 4X^2G_{4XXX}\right) \\
&-12XH^2 \left(G_{5X}+5XG_{5\phi X} +2X^2G_{5\phi XX}\right) \\
&+14\dot\phi H^3\left(3G_{5X} +7XG_{5XX} +2X^2G_{5XXX} \right)\, , \\
HM_*^2\alpha_\mathrm{B} & \equiv 2\dot\phi\left(XG_{3X} -G_{4\phi} -2XG_{4\phi X} \right)\\
& +8XH\left(G_{4X}+2XG_{4XX} -G_{5\phi}-XG_{5\phi x}\right) \\
& +2\dot\phi XH^2\left(3G_{5X} +2XG_{5XX} \right), \\
M_*^2\alpha_\mathrm{T} & \equiv 2X\left(2G_{4X} -2G_{5\phi} -(\ddot{\phi}-\dot\phi H)G_{5X}\right).
\end{split}
\end{equation}
Here $M_*$ is the Planck mass and $\alpha_\mathrm{M}$ describes its logarithmic time evolution. $\alpha_\mathrm{K}$ describes the kinetic energy and will largely be unconstrained by observations of the LSS \citep{alonso_observational_2017,spurio_mancini_testing_2018,reischke_investigating_2019}.
In contrast, the braiding $\alpha_\mathrm{B}$ describes how $\phi$ mixes with the scalar perturbations of the metric. Lastly $\alpha_\mathrm{T}$ basically describes the propagation speed of tensorial modes and how it differs from normal null geodesics and has been constrained to be very close to zero \citep{abbott_gw170817:_2017,abbott_multi-messenger_2017} and will therefore be safely ignored from any LSS analysis. 
In principle, the remaining functions are completely free, however, a common choice for a parametrisation would be
\begin{equation}\label{eq:parametrization_de}
\alpha_i = \hat{\alpha}_i \Omega_\mathrm{DE} + c_i,
\end{equation}
since in such a way the modifications track the accelerated expansion of the Universe. For more details we refer to  \citet{linder_is_2016} and \citet{alonso_observational_2017,gleyzes_unifying_2014}. As remarked, we will focus on $\hat{\alpha}_B$ and $\hat{\alpha}_M$. Since $\hat{\alpha}_M$ affects the propagation of gravitational waves as well (mainly by a damping term) it can also be constrained from more local experiments. We refer the read to \citet{ezquiaga_dark_2017} for viable regions in the Horndeski space after GW170817. With the remaining free parameters Horndeski theories still include quintessence, $f(R)$ or Brans-Dicke.

The solutions to the linear perturbation equations is provided by \texttt{HiClass} \citep{zumalacarregui_hiclass:_2016}, an extension to the \texttt{Class} code \citep{blas_cosmic_2011, lesgourgues_cosmic_2011} for Horndeski theories. From the code linear power spectra of the Bardeen potentials and the matter density are readily available.

\begin{table}
    \centering
    \begin{tabular}{ccc}
     parameter &  fiducial value & interpretation\\
     \hline\hline 
        $f_\mathrm{sky}$ & 0.15 & sky fraction \\
        $\bar{n}$ & 30 arcmin$^{-2}$ & source density \\
        $n_\mathrm{bin}$ & 6 & tomographic bins\\
        $\sigma_\epsilon$  &0.3 & ellipticity dispersion\\
        $f_\mathrm{red}$ & 0.3 & elliptical fraction \\
        $(\beta,\;z_0)$ & $(1.5,\;0.9)$ & redshift distribution \\
                $(\ell_\mathrm{min},\;\ell_\mathrm{max})$ & $(10,\;300)$ & multipole range \\
        $D/(3\times 10^{-5})$ & 1 & alignment strength \\ 
        $\hat\alpha_B$ & 0.05 & braiding \\
        $\hat\alpha_M$ & 0.05 & Planck mass running \\
    \end{tabular}
    \caption{Survey settings and the fiducial parameters with description. All other cosmological parameters are set to the best fit values as measured by \citet{planck_collaboration_planck_2018}.}
    \label{tab:table_1}
\end{table}

\section{Probing gravitational slip with shape correlations}\label{sect_slip}
In this section we will briefly discuss the two effects altering the shapes of background galaxies: gravitational lensing and intrinsic alignments.

\subsection{Weak gravitational lensing}
Bundle of light rays travelling from distant sources are distorted due to varying gravitational potentials of the LSS. Due to the null condition of photon geodesics lensing measures $\Phi$ and $\Psi$ \citep[up to a sign, depending on the signature of the metric,][]{acquaviva_weak_2004}. The lensing potential is the line-of-sight projection of these two quantities
\begin{equation}\label{eq:lensingpot}
\psi_i = \int_0^{\chi_H}\mathrm{d}\chi W_{i}(\chi) (\Phi +\Psi)\;,
\end{equation}
where $\chi_H=c/H$ and $W_i(\chi)$ is a weight function:
\begin{equation}
W_{i}(\chi) = \frac{G_i(\chi)}{a\chi}\;.
\end{equation}
The lensing efficiency function is given by
\begin{equation}
G_i(\chi) = 
\int _{\mathrm{min}(\chi,\chi_i)}^{\chi_{i+1}}\mathrm{d}\chi'p(\chi')\frac{\mathrm{d}z}{\mathrm{d}\chi'}\left(1-\frac{\chi}{\chi'}\right)\;.
\end{equation}
The distribution $p(z)\mathrm{d}z$ of sources in redshift $z$ takes the usual form for a flux limited survey \citep{laureijs_euclid_2011}
\begin{equation}
p(z)\mathrm{d}z \propto z^2\exp\left[-\left(\frac{z}{z_0}\right)^\beta\right]\;.
\end{equation}
We choose $z_0 = 0.9 $ and $\beta = 3/2$. Lastly the index $i$ labels the tomographic bin of the source sample for which we choose 6 equally populated bins which has been shown to be close to optimal for most cosmological parameters \citep{sipp_optimising_2020}.
With these ingredients, the angular power spectrum of the lensing potential $\psi$ in tomographic bins $i$ and $j$ is given by
\begin{equation}\label{eq:cosmic_shear_power}
C_{\psi_i\psi_j}(\ell) = 
\int_0^{\chi_H} \frac{\mathrm{d}\chi}{\chi^2}W_{\psi_i}(\chi)W_{\psi_j}(\chi)P_{\Phi+\Psi}(\ell'/\chi,\chi)\;,
\end{equation}
where we defined the power spectrum of the the sum of the two Bardeen potentials $\Phi$ and $\Psi$. It should be noted that \cref{eq:cosmic_shear_power} uses the Limber approximation. For cosmic shear, however, this expression is very accurate up to large angular scales due to the broad lensing kernel $W_i(\chi)$. The corresponding converge, $\kappa$, (shear) spectra is then obtained via the relation $\Delta\Psi = 2\kappa$, which amounts to a factor of $\ell^4/4$ in harmonic space for the spectrum. Observed lensing spectra are noisy due to Poissonian shape noise:
\begin{equation}
\hat{C}_{\kappa_i\kappa_j}(\ell) = \frac{\ell^4}{4}C_{\psi_i\psi_j} + \sigma_\epsilon^2\frac{n_\mathrm{bin}}{\bar{n}}\delta_{ij}.
\end{equation}
with the intrinsic ellipticity dispersion $\sigma_\epsilon = 0.3$, the number of tomographic bins $n_\mathrm{bin}$ and the mean number density of galaxies $\bar{n} = 30 \;\mathrm{arcmin}^{-2}$.

\subsection{Tidal alignment of elliptical galaxies}
The alignment models are both based on tidal interaction of galaxies with the LSS and we refer the reader to \citet{tugendhat_angular_2017} for more details. In this section we will just provide the basic definitions and discuss the impact of gravitational slip on intrinsic alignments. We require correlations of the tidal shear:
\begin{equation}
\begin{split}
C_{\alpha\beta\gamma\delta}(r)\equiv & \ \bra\Phi_{\alpha\beta}(\bmath{x})\Phi_{\gamma\delta}(\bmath{x}^\prime)\ket\\
= & \
(\delta_{\alpha\beta}\delta_{\gamma\delta}+\delta_{\alpha\gamma}\delta_{\beta\delta}+\delta_{\alpha\delta}\delta_{\beta,\gamma})\zeta_2(r)+\\ 
&(\hat{r}_\alpha \hat{r}_\beta \delta_{\gamma\delta}+\mathrm{5~perm.}) \zeta_3(r)+\\
&\hat{r}_\alpha \hat{r}_\beta \hat{r}_\gamma \hat{r}_\delta \zeta_4(r),
\end{split}
\end{equation} 
where $r=\left|\bmath{x}-\bmath{x}^\prime\right|$ and $\zeta_n(r)$ is, 
\label{sec:quadalign}
\begin{equation}
\zeta_n(r) = \left(-1\right)^n r^{n-4}\int\frac{\mathrm{d}{k}}{2\pi^2}\:P_\Phi(k)\,k^{n+2}\,j_n(kr),
\label{eq:zeta_n}
\end{equation}
as Fourier-transforms the spectrum $P_\Phi(k)$ and its derivatives under the assumption of isotropy \citep{crittenden_spin-induced_2001}. It should be noted that in contrast to \cref{eq:cosmic_shear_power} only the power spectrum of $\Phi$, i.e. the time-time component arises.

The picture of an elliptical galaxy is that it is described by a virialised system in which the stars perform random motion with a constant velocity dispersion $\sigma^2$ and anisotropy parameter $\beta$ provided a gravitational potential $\Phi$. The density of the cloud of stars follows from the Jeans-equation $\rho\propto\exp(-\Phi/\sigma^2)$ if $\beta = 0$ Perturbing the isotropic system slightly by adding a quadrupole to the potential the, the density of particles would change to
\begin{equation}
\rho \propto 
\exp\left(-\frac{\Phi(\bmath{x})}{\sigma^2}\right)
\times
\left(1-\frac{1}{2\sigma^2}\frac{\partial^2\Phi(\bmath{x}_0)}{\partial x_\alpha\partial x_\beta}x_\alpha x_\beta\right),
\label{eqn_jeans}
\end{equation}
where $\boldsymbol{x}_0$ is the peak of the original density profile. This in turn produces an ellipticity linear in the tidal field with a complex ellipticity $\epsilon = \epsilon_+ + \mathrm{i}\epsilon_\times$ given by
\begin{equation}
\epsilon = D \left(\frac{\partial^2\Phi}{\partial x^2} - \frac{\partial^2\Phi}{\partial y^2} + 2\mathrm{i} \frac{\partial^2\Phi}{\partial x\partial y}\right),
\label{eq:linearmodel}
\end{equation}
assuming the sky plane to coincide with the $x,y$-plane. Here $D$ quantifies the response of the ellipticity to the tidal field.
The correlations are given by
\begin{align}
\left\langle\epsilon_+ \, \epsilon_+'\right\rangle (\boldsymbol{r})&= D^2\,\left(4\,\zeta_2(r)+4\,\sin^2(\alpha)\,\zeta_3(r)+\sin^4(\alpha)\,\zeta_4(r)\right),\\
\label{eq:C3Dpp}
\left\langle\epsilon_{\times} \, \epsilon_{\times}'\right\rangle(\boldsymbol{r}) &= 4\,D^2\,\left(\zeta_2(r)+\sin^2(\alpha)\,\zeta_3(r)\right).
\end{align}
Making extensive use of the Limber projection in real space \citep{limber_analysis_1954, loverde_extended_2008}, one finds
\begin{equation}
\left\langle\epsilon_{a,i} \, \epsilon_{a,i}'\right\rangle(\theta) = 
\int_{0}^{\chi_H}\mathrm{d}\chi\: n_i(\chi)\:
\int_{\chi_{i}}^{\chi_{i+1}}\mathrm{d}\chi'\: n_{i}(\chi') 
\left\langle\epsilon_{a} \, \epsilon_{a}'\right\rangle\Big(\boldsymbol{r}(\chi, \chi')\Big)
\end{equation}
where $a \in \left\{+,\times\right\}$.
With 
\begin{equation}
C^{\epsilon,\text{II}}_{\pm,i}(\theta) = \left\langle\epsilon_{+,i} \, \epsilon_{+,i}'\right\rangle \pm \left\langle\epsilon_{\times,i} \, \epsilon_{\times,i}'\right\rangle,
\label{eq:Cpmi}
\end{equation}
one can calculate the ellipticity $E$- and $B$-mode spectra of linear alignment $C_{i}^{\epsilon}(\ell)$ via a Fourier transform,
\begin{align}
\label{eq:ebmode}
C^{\epsilon,\text{II}}_{{E},i}(\ell)&=\pi \int \theta \mathrm{d} \theta ~\Big(C^{\epsilon,\text{II}}_{+,i} (\theta) \, J_0(\ell\theta)+C^{\epsilon,\text{II}}_{-,i}(\theta)J_4(\ell\theta) \Big)\;,\\
\label{eq:ebbmode}
C^{\epsilon,\text{II}}_{{B},i}(\ell)&=\pi \int \theta\mathrm{d} \theta ~\Big(C^{\epsilon,\text{II}}_{+,i} (\theta) \, J_0(\ell\theta)-C^{\epsilon,\text{II}}_{-,i}(\theta)J_4(\ell\theta) \Big)\;,
\end{align}
where $J_\nu(x)$ are regular Bessel functions.

Even more important than the intrinsic alignment auto-correlation (II) is the GI-correlations which are cross-correlations between weak lensing and intrinsic alignments since we then probe all possible combinations of $\Phi$ and $\Psi$.
For the model considered here, the 3d correlation functions assume a familiar form
\begin{equation}
\left\langle\gamma_+ \, \epsilon_+'\right\rangle (\boldsymbol{r}) = -\int_{0}^{\chi_H}\mathrm{d} \chi ~ \frac{\tilde W(\chi)}{D}\,\left\langle\epsilon_+ \, \epsilon_+'\right\rangle (\boldsymbol{r}),\\
\end{equation}
and
\begin{equation}
\left\langle\gamma_\times \, \epsilon_\times'\right\rangle (\boldsymbol{r}) =
\int_{0}^{\chi_H} \mathrm{d} \chi \frac{\tilde W(\chi)}{D}\,\left\langle\epsilon_\times \, \epsilon_\times'\right\rangle (\boldsymbol{r}),\\
\end{equation}
where $\tilde W(\chi) = (1+ \Psi/\Phi) W(\chi)$ to account for the proper Weyl potential for the lensing part.
In complete analogy to the II-case:
\begin{equation}
\left\langle\gamma_{a,i} \, \epsilon_{a,j}'\right\rangle(\theta) =
\mp\int_{0}^{\chi_H} \hspace{-2mm}\mathrm{d} \chi \int_{\chi_{j}}^{\chi_{j+1}} \hspace{-2mm}\mathrm{d} \chi'\frac{\tilde W_{i}(\chi)\,n_{j}(\chi')}{D}\,\left\langle\epsilon_a \, \epsilon_a'\right\rangle\Big(\boldsymbol{r}(\chi, \chi')\Big)
.
\label{eq:GI2dcorr}
\end{equation}
The $E$- and $B$-mode spectra are then readily calculated via \cref{eq:ebmode,eq:ebbmode}.

\begin{table}
    \centering
    \begin{tabular}{cccc}
     type &  $1\sigma$ error ($\hat\alpha_B$) & $1\sigma$ error ($\hat\alpha_M$) & 1$\sigma$ error $D$\\
     \hline\hline 
        $GG$ &  0.12  & 0.13 & - \\
        II+GI & 0.32 & 0.30 & - \\
        GG+II+GI & 0.11 & 0.10 & - \\
            $D$ marginalized      & 0.11 & 0.12 & 0.022
    \end{tabular}
    \caption{Constraints on the modified gravity parameters and the alignment parameter for different settings.}
    \label{tab:table_2}
\end{table}

\section{Results}\label{sect_results}
We forecast the possible constraints on the two remaining degrees of freedom $\hat{\alpha}_M$ and $\hat{\alpha}_B$ by means of a Fisher-analysis, where the likelihood is constructed from the angular ellipticity spectra with a Gaussian-approximated covariance containing cosmic variance and a shape noise term. The strength of intrinsic shape correlations in the cosmic shear survey will largely affect the constraints and it depends several aspects.

\begin{figure}
    \centering
    \includegraphics[width = 0.45\textwidth]{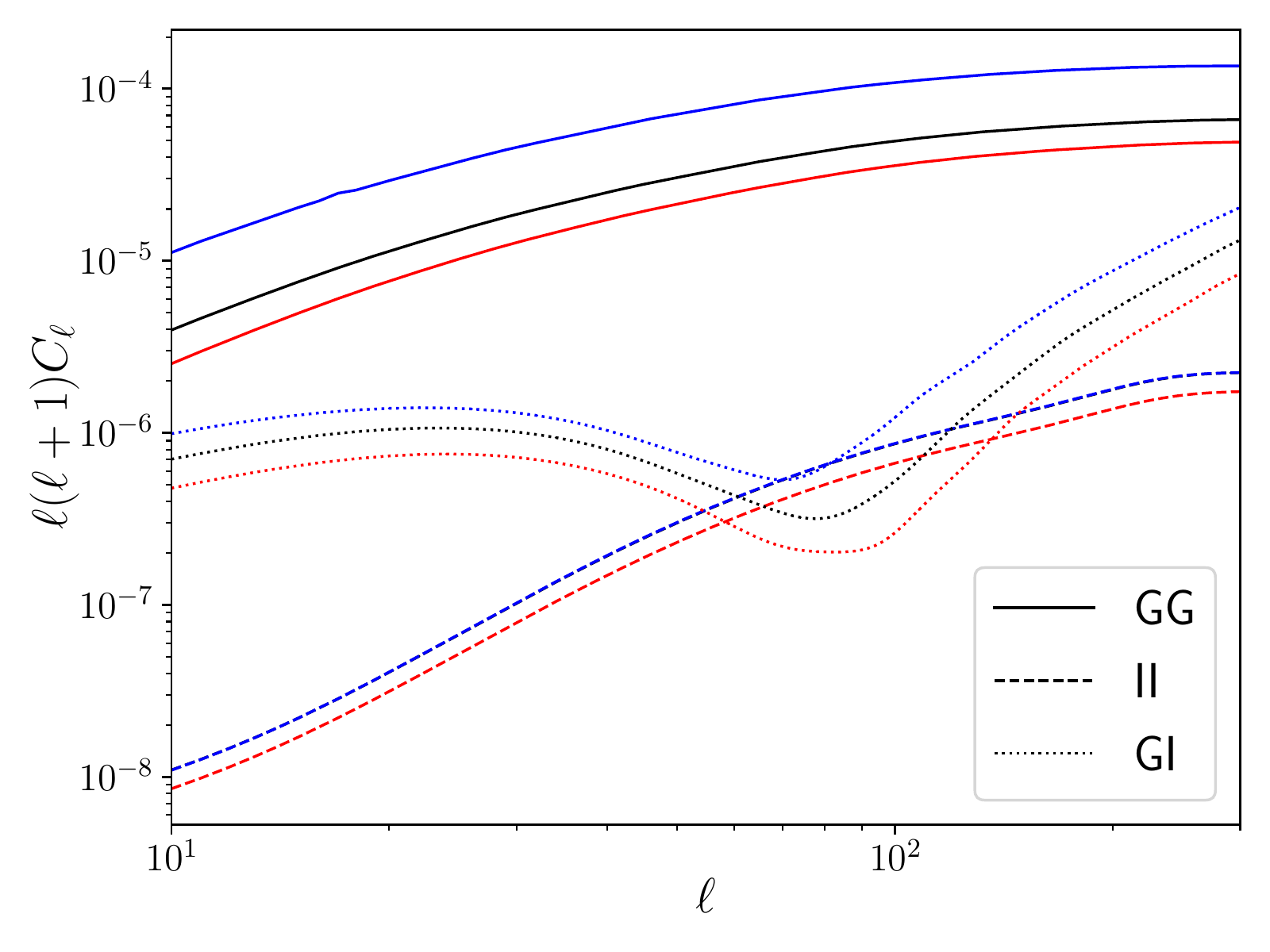}
    \caption{Tomographic ellipticity spectra. The black lines correspond to the a $\Lambda$CDM cosmology while the red and blue line are Horndeski theories with $\hat{\alpha}_M = 2.5$ and  $\hat{\alpha}_B = 2.5$ respectively. Different line styles depict the pure lensing signal (GG), pure alignment signal (II) and the cross-correlation (GI). We assume two tomographic bins in this case. To avoid any clutter, the $GG$ and $II$ contribution are shown for the auto-correlation of the first tomographic bin, while the $GI$ contributions correlates the second and the first bin.}
    \label{fig:spectra}
\end{figure}

Most obviously, the value of the alignment parameter $D$ is not yet well known and there is no unanimity on its numerical value. The parameter determines the relative importance of the $GI$- and $II$-terms in the angular ellipticity correlations, and introducing proportionalities of the signal with $D$ and $D^2$, respectively. The differences between the three contributions to the ellipticity correlations are, due to their different scaling with $D$, non-degenerate to some extent. 

\citet{heymans_cfhtlens_2013} for example measured a significant alignment signal from CFHTLenS which was used by \citet{tugendhat_angular_2017} to determine $D = 9.5\times 10^{-5}c^2$, whereas \citet{hilbert_intrinsic_2016-1} measured $D = 1.5\times 10^{-4}c^2$ directly from simulations. Instead \citet{zjupa_intrinsic_2020} found a very small alignment amplitude (over a magnitude smaller) in the V-band of \texttt{IllustrisTNG} depending both on redshift \citep[a trend also observed in][]{tenneti_intrinsic_2015,samuroff_advances_2020} and environment \citep{reischke_environmental_2019}. It also depends on the smoothing scale which is applied to the tidal field. This is, however, by definition degenerate with the amplitude. In summary, the alignment strength $D$ is still up for debate and we will choose a compromise between the different values in the literature here to be conservative. Stronger alignments would in fact facilitate the measurement.

The second aspect determining the overall contribution from intrinsic ellipticity correlations is the fraction of the galaxy type with a certain alignment mechanism. As discussed in \cref{sec:intro} we will consider alignments only for luminous red (elliptical) galaxies. Stage IV survey, however, will observe a large fraction of blue galaxies. In order to account for this intrinsic ellipticities pick up a factor $f_\mathrm{red}$ describing the fraction of elliptical galaxies in the survey
\begin{equation}
C^{\epsilon,\text{II}}_{{E},i}(\ell) \to f^2_\mathrm{red}C^{\epsilon,\text{II}}_{{E},i}(\ell)\;, \quad 
C^{\epsilon,\text{GI}}_{{E},i}(\ell) \to   f_\mathrm{red}C^{\epsilon,\text{GI}}_{{E},i}(\ell)\;.
\end{equation}
All parameters we assume for our forecast are summarised in \cref{tab:table_1}. As shown by \citep{2020arXiv200504604G}, there is no particular advantage in pre-selecting elliptical galaxies for obtaining a clean sample, where all galaxies contribute to intrinsic shape correlations, in contrast to the full sample, where only the fraction $f_\mathrm{red}$ and $f_\mathrm{red}^2$ contribute to the $GI$ and $II$-type correlations. In this trade-off, the signal would be weaker by these factors, but this is largely compensated by the overall smaller shape noise terms.

\Cref{fig:spectra} shows the signal we are looking for: The black lines show the $\Lambda$CDM (all $\hat{\alpha}_i = 0$) case while blue and red have $\hat{\alpha}_B$ and $\hat{\alpha}_B$ changed to 2.5 respectively. The solid lines show the lensing spectra, the dotted line the $GI$ contribution and the dashed line the $II$ spectrum. If we compare the relative amplitude of the spectra we see that the $GI$ spectrum is dominating the $II$ contribution more than for example in \citet{tugendhat_angular_2017}. This is a result of the reduced alignment parameter $D$. For definiteness we assumed two tomographic bins and show only the auto-correlation of the first bin for $GG$ and $II$ and the cross-correlation in case of GI. It should also be noted that the $GI$ contribution is negative which has an twofold consequences for the inference process: it removes signal from the ellipticity correlations but at the same time it makes tomographic bins more independent since the $II$ contribution is positive. Considering the dependence on the Horndeski parameters, we see that the $II$ contribution is largely unaffected by the braiding, $\hat{\alpha}_B$, which is expected since this mainly changes the gravitational slip. In contrast it is affected by the running of the Planck mass, $\hat{\alpha}_M$, as this directly changes the Poisson equation and therefore the tidal field of $\Phi$. The situation is different for $GI$ which is sensitive to both parameters via the Weyl potential from lensing and $\Phi$ from the tidal alignment. 

\begin{figure}
    \centering
    \includegraphics[width = 0.45\textwidth]{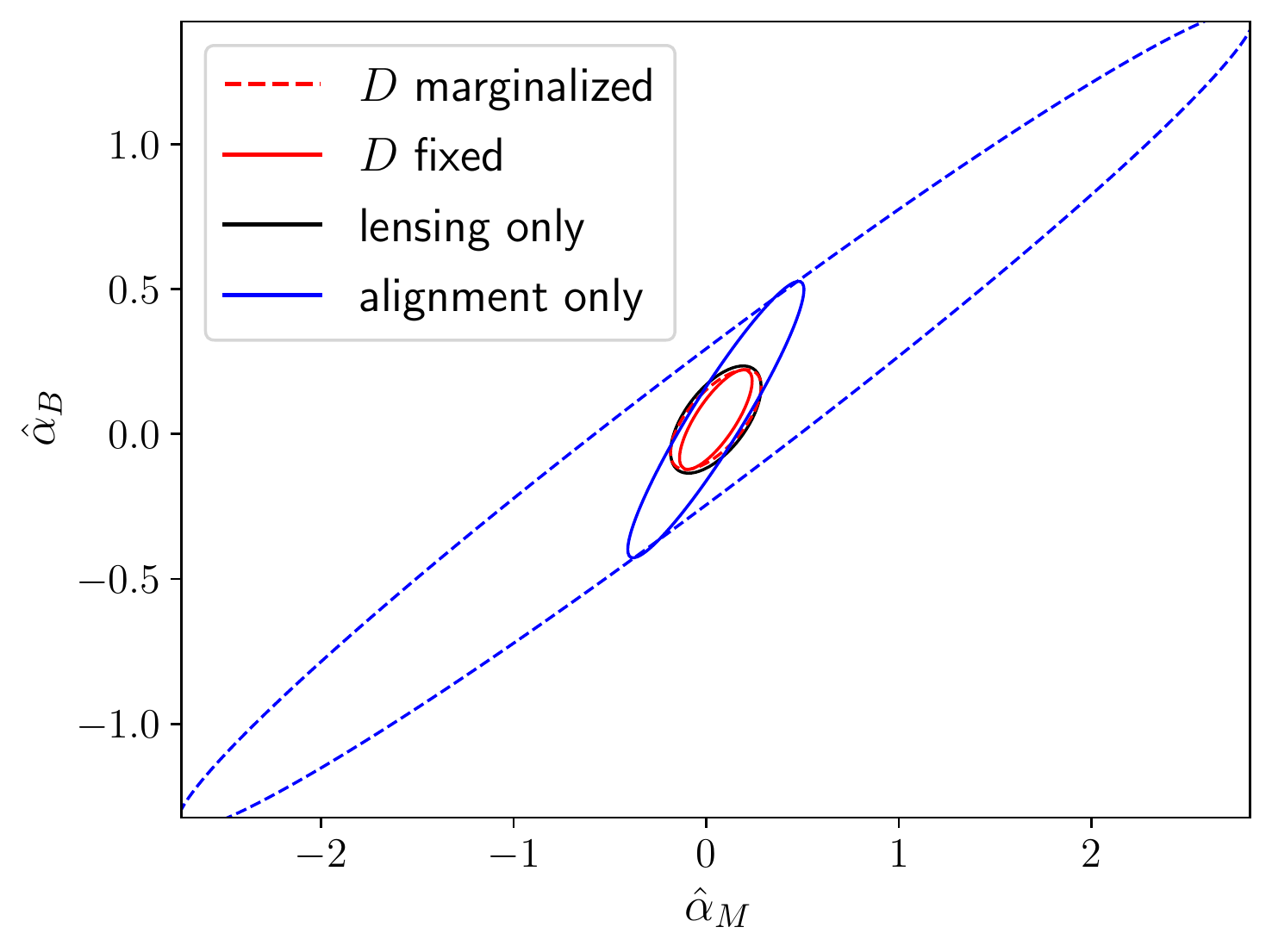}
    \caption{One sigma constraints for a \textit{Euclid}-like survey. Other cosmological parameters have been set to their fiducial values and we only vary $\hat{\alpha}_M$ and  $\hat{\alpha}_B$. The black and blue ellipse correspond to lensing and alignment only respectively. In red we show the combination of the two. Dashed ellipses show the situation when we marginalize over the alignment amplitude $D$.}
    \label{fig:fisher}
\end{figure}

We now assume a survey as described in \cref{tab:table_1} to forecast constraints on $\hat{\alpha}_B$ and $\hat{\alpha}_M$. To this end we assume Gaussian statistics and perform a Fisher forecast \citep[see e.g.][]{tegmark_karhunen-loeve_1997-1}. We summarize the results in \cref{tab:table_2}. \Cref{fig:fisher} shows the Cram\'{e}r-Rao ellipse for the two Horndeski parameters for the cases considered here. Solid lines show constraints where the alignment amplitude (if necessary) is fixed to its fiducial value while dashed line marginalize over it. Furthermore we consider the case where we only use lensing (black) only alignment (blue) and the combination of the two (red). It should be noted that we include the $GI$ contribution in the alignment only case. Since measuring the ellipticities is not a combination of probes but rather a single measurement, this split is of course somewhat arbitrary since the observed correlation will always include contributions. Finally, we fix all other cosmological parameters to their fiducial values \citep{planck_collaboration_planck_2018} and only fit the modified gravity and alignment parameters. One can see IA adding roughly 10 or 30 per cent to the precision on $\hat{\alpha}_B$ and $\hat{\alpha}_M$ respectively. The stronger sensitivity on $\hat{\alpha}_M$ again stems from the fact that the $II$ contribution is not particularly sensitive on $\hat{\alpha}_B$ as discussed before. When marginalising over the alignment amplitude $D$ the additional constraints from intrinsic alignments are weak and they are not able to deliver any benefit over lensing alone. The reason for this can already be seen in \cref{fig:spectra}: The dependence on $\hat{\alpha}_M$ is similar to a different alignment amplitude. We also tested the case where each tomographic bin requires an independent alignment amplitude $D_i$, $i=1,...,n_\mathrm{bin}$. This made almost no difference on the constraints on modified gravity mainly because as soon as the alignment parameter is left free already all of the signal from $GI$ and $II$ needs to be sacrificed for fitting $D$. It would therefore be necessary to provide external information on $D$, either by other observations or from numerical simulations.

Although we compute the observable ellipticity correlations for Horndeski-gravity as a specific model of modified gravity currently heavily discussed in the literature, we would like to point out that one can make use of their constraining power if the alignment parameter $D$ is provided from other sources, like simulations or from observations in the local Universe with reconstructed tidal gravitational fields. Both would be viable as the reaction of an elliptical galaxy to a tidal field is governed by Newtonian physics valid on the scale of a galaxy, in the mainstream concept assuming dark matter as the dominating component of matter, and Newtonian physics being necessarily the limit of any viable theory of modified gravity. Deviations in the gravitational interaction would then be measured relative to that, in our case by gravitational light deflection. The contribution of the gravitational, Newtonian potential as the first Bardeen-potential on these scales would then be completed by a possibly modified curvature perturbation as the second Bardeen-potential.

If one has alternative access to the alignment parameter, even in a statistical way as a prior to the inference, the constraining power of intrinsic alignments can be used even for a standard $\Lambda$CDM or $w$CDM cosmology. But it is in combination with gravitational lensing through the independent measurement of the Bardeen-potential that the unique sensitivity to modified gravity comes to bear. In the simplest, empirical application one could determine the gravitational slip parameter \citep{2009PhRvD..80b3532D}, or the parameters $\eta$ and $\mu$ for a modified Poisson-equation \citep{amendola_cosmology_2012}. On a more fundamental level, our test would be sensitive to screening mechanisms \citep{falck_vainshtein_2014} or effects of anisotropic stress within general relativity \citep{majerotto_probing_2012}, which likewise generates differences in the Bardeen potentials.

In comparison to the established and conventional method of combining lensing and the clustering or the peculiar motion of galaxies \citep{heavens_nonlinear_1998, fosalba_probing_2007, mortonson_dark_2013, bellini_constraints_2016}, which take on the role of nonrelativistic test particles in this case, we would like to point out that both gravitational lensing and intrinsic alignments are both sensitive to tidal gravitational fields, so they measure the spectrum of fluctuations naturally on the same scale, unlike velocity fields, which provide information on larger scales as first derivatives of the gravitational potential. Furthermore, IAs and lensing manifest themselves as ellipticity correlations and are therefore derived from the same data set. It would be interesting to see to what extend the complementarity of intrinsic alignments as a source of cosmological information can be used, similar to the well-established complementarity between lensing and velocities \citep{shapiro_complementarity_2012, kim_complementarity_2019}. 

\section{Summary}\label{sect_summary}
In this paper we investigated the impact of modified gravity theories of the Horndeski class on intrinsic ellipticity correlations and the resulting consequences for inference with cosmic shear data from stage IV surveys. Our motivation was to turn intrinsic alignments from a nuisance to a cosmological probe. We use their unique sensitivity to a single  Bardeen-potential (the gravitational potential $\Phi$) , and contrast them to gravitational lensing, which depends on both Bardeen-potentials (the sum of the gravitational potential $\Phi$ and the curvature perturbation $\Psi$), to carry out a cosmological test of gravity. While we have focused on the impact of inference of modified gravity and in particular the Horndeski class, similar conclusions can be drawn for other cosmological parameters, for phenomenological parameterisations of the Poisson-equation, or measurements of anisotropic stress within standard general relativity. 

The analysis is restricted to large scales $\ell < 300$ where we can assume that the alignment of red (elliptical) galaxies can be described by a simple model, linear in the tidal shear, and where for a typical survey the density field can be considered in a stage of evolution well described by linear theory. For blue (spiral) galaxies the alignment is more short ranged, since alignment models are at least quadratic in the tidal shear, and the restriction to low multipoles in fact discards any contribution of spiral alignments, for which there is only inconclusive evidence from simulations.

We summarize our findings as follows:
\begin{enumerate}
\item{IAs carry in principle complementary about the metric perturbations and can therefore help to break degeneracies when measuring the gravitational slip. In particular in the context of Horndeski gravity, the $II$ component will mainly be sensitive to changes in the Planck mass over cosmic time.}

\item{When the alignment strength is know a prior (for example through simulations), IAs can provide a marked increase in sensitivity (of modified gravity parameters) when treated as a model prediction instead of as a nuisance.  This statement strongly depends on the overall coupling strength of the intrinsic ellipticity to the tidal field which is still debated in the literature (see \cref{sect_results} for a discussion). The results presented here are for a conservative value of the coupling strength.}

\item{If the data itself has to measure the coupling strength the whole alignment signal is used to fit the alignment parameter $D$, leaving nearly no signal-to-noise for determining additional parameters. In this case, the constraints on $\hat{\alpha}_B$ and $\hat{\alpha}_M$ are completely dominated by the lensing signal. Since the important quantity is the relative sensitivity between lensing and IAs one can suspect very similar outcomes for other parameters: It is difficult to imagine how this can be circumvented, as the alignment parameter $D$ determines the amplitude of the IA signal in a scale independent way.}
\end{enumerate}

With the above conclusions it is evident that IAs will stay a nuisance which needs to be accounted for in future weak lensing surveys and can only add significant additional information if their strength is measured in simulations or provided by other means: Even stage IV surveys will not possess enough statistical power to determine the alignment parameter $D$ independently from the two $\hat\alpha$-parameters needed for Horndeski gravity. Improving this situation requires a more refined for spiral galaxies since the tidal torquing model does not seem to be well produced by numerical simulations \citep{zjupa_intrinsic_2020}; instead one found empirical evidence for a linear alignment model, whose mechanisms are yet to be determined. One could imagine that spiral galaxies would add statistical power to the measurement due to the large fraction of blue galaxies in future surveys.

\vspace{2mm}

{\bf Data Availability}: The data underlying this article will be shared on reasonable request to the corresponding author.

\section*{Acknowledgements}
RR is supported by the European Research Council (Grant No. 770935). The work of VB is funded by a fellowship from ``la Caixa'' Foundation (ID 100010434). The fellowship code is LCF/BQ/DI19/11730063.

\bibliographystyle{mnras}
\bibliography{MyLibrary}

\begin{thebibliography}{}
\makeatletter
\relax
\def\mn@urlcharsother{\let\do\@makeother \do\$\do\&\do\#\do\^\do\_\do\%\do\~}
\def\mn@doi{\begingroup\mn@urlcharsother \@ifnextchar [ {\mn@doi@}
  {\mn@doi@[]}}
\def\mn@doi@[#1]#2{\def\@tempa{#1}\ifx\@tempa\@empty \href
  {http://dx.doi.org/#2} {doi:#2}\else \href {http://dx.doi.org/#2} {#1}\fi
  \endgroup}
\def\mn@eprint#1#2{\mn@eprint@#1:#2::\@nil}
\def\mn@eprint@arXiv#1{\href {http://arxiv.org/abs/#1} {{\tt arXiv:#1}}}
\def\mn@eprint@dblp#1{\href {http://dblp.uni-trier.de/rec/bibtex/#1.xml}
  {dblp:#1}}
\def\mn@eprint@#1:#2:#3:#4\@nil{\def\@tempa {#1}\def\@tempb {#2}\def\@tempc
  {#3}\ifx \@tempc \@empty \let \@tempc \@tempb \let \@tempb \@tempa \fi \ifx
  \@tempb \@empty \def\@tempb {arXiv}\fi \@ifundefined
  {mn@eprint@\@tempb}{\@tempb:\@tempc}{\expandafter \expandafter \csname
  mn@eprint@\@tempb\endcsname \expandafter{\@tempc}}}

\bibitem[\protect\citeauthoryear{Abbott et~al.,}{Abbott
  et~al.}{2017a}]{abbott_gw170817:_2017}
Abbott B.~P.,  et~al., 2017a, \mn@doi [Physical Review Letters]
  {10.1103/PhysRevLett.119.161101}, 119, 161101

\bibitem[\protect\citeauthoryear{Abbott et~al.,}{Abbott
  et~al.}{2017b}]{abbott_multi-messenger_2017}
Abbott B.~P.,  et~al., 2017b, \mn@doi [\apjl] {10.3847/2041-8213/aa91c9}, 848,
  L12

\bibitem[\protect\citeauthoryear{Acquaviva, Baccigalupi  \& Perrotta}{Acquaviva
  et~al.}{2004}]{acquaviva_weak_2004}
Acquaviva V.,  Baccigalupi C.,   Perrotta F.,  2004, \mn@doi [\prd]
  {10.1103/PhysRevD.70.023515}, 70, 023515

\bibitem[\protect\citeauthoryear{Albrecht et~al.,}{Albrecht
  et~al.}{2006}]{albrecht_report_2006}
Albrecht A.,  et~al., 2006, preprint

\bibitem[\protect\citeauthoryear{Alonso, Bellini, Ferreira  \&
  Zumalac{\'a}rregui}{Alonso et~al.}{2017}]{alonso_observational_2017}
Alonso D.,  Bellini E.,  Ferreira P.~G.,   Zumalac{\'a}rregui M.,  2017,
  \mn@doi [\prd] {10.1103/PhysRevD.95.063502}, 95, 063502

\bibitem[\protect\citeauthoryear{Amendola et~al.,}{Amendola
  et~al.}{2012}]{amendola_cosmology_2012}
Amendola L.,  et~al., 2012, ArXiv e-prints, 1206.1225

\bibitem[\protect\citeauthoryear{Baker, Bellini, Ferreira, Lagos, Noller  \&
  Sawicki}{Baker et~al.}{2017}]{baker_strong_2017}
Baker T.,  Bellini E.,  Ferreira P.~G.,  Lagos M.,  Noller J.,   Sawicki I.,
  2017, \mn@doi [Physical Review Letters] {10.1103/PhysRevLett.119.251301},
  119, 251301

\bibitem[\protect\citeauthoryear{Bartelmann \& Schneider}{Bartelmann \&
  Schneider}{2001}]{bartelmann_weak_2001}
Bartelmann M.,  Schneider P.,  2001, \mn@doi [\physrep]
  {10.1016/S0370-1573(00)00082-X}, 340, 291

\bibitem[\protect\citeauthoryear{Battye \& Pearson}{Battye \&
  Pearson}{2013}]{battye_parametrizing_2013}
Battye R.~A.,  Pearson J.~A.,  2013, \mn@doi [\prd]
  {10.1103/PhysRevD.88.061301}, 88, 061301

\bibitem[\protect\citeauthoryear{Bellini \& Sawicki}{Bellini \&
  Sawicki}{2014}]{bellini_maximal_2014}
Bellini E.,  Sawicki I.,  2014, \mn@doi [\jcap]
  {10.1088/1475-7516/2014/07/050}, 7, 050

\bibitem[\protect\citeauthoryear{Bellini, Cuesta, Jimenez  \& Verde}{Bellini
  et~al.}{2016}]{bellini_constraints_2016}
Bellini E.,  Cuesta A.~J.,  Jimenez R.,   Verde L.,  2016, \mn@doi [JCAP]
  {10.1088/1475-7516/2016/02/053}, 1602, 053

\bibitem[\protect\citeauthoryear{Berti et~al.,}{Berti
  et~al.}{2015}]{berti_testing_2015}
Berti E.,  et~al., 2015, \mn@doi [Classical and Quantum Gravity]
  {10.1088/0264-9381/32/24/243001}, 32, 243001

\bibitem[\protect\citeauthoryear{Bertschinger \& Zukin}{Bertschinger \&
  Zukin}{2008}]{bertschinger_distinguishing_2008}
Bertschinger E.,  Zukin P.,  2008, \mn@doi [Phys.Rev.]
  {10.1103/PhysRevD.78.024015}, D78, 024015

\bibitem[\protect\citeauthoryear{Beutler et~al.,}{Beutler
  et~al.}{2017}]{beutler_clustering_2017}
Beutler F.,  et~al., 2017, \mn@doi [Mon. Not. R. Astron. Soc.]
  {10.1093/mnras/stw2373}, 464, 3409

\bibitem[\protect\citeauthoryear{Bhattacharya, Nagai, Shaw, Crawford  \&
  Holder}{Bhattacharya et~al.}{2012}]{bhattacharya_bispectrum_2012}
Bhattacharya S.,  Nagai D.,  Shaw L.,  Crawford T.,   Holder G.~P.,  2012,
  \mn@doi [\apj] {10.1088/0004-637X/760/1/5}, 760, 5

\bibitem[\protect\citeauthoryear{Blas, Lesgourgues  \& Tram}{Blas
  et~al.}{2011}]{blas_cosmic_2011}
Blas D.,  Lesgourgues J.,   Tram T.,  2011, \mn@doi [Journal of Cosmology and
  Astroparticle Physics] {10.1088/1475-7516/2011/07/034}, 07, 034

\bibitem[\protect\citeauthoryear{Blazek, McQuinn  \& Seljak}{Blazek
  et~al.}{2011}]{blazek_testing_2011}
Blazek J.,  McQuinn M.,   Seljak U.,  2011, Journal of Cosmology and
  Astroparticle Physics, 2011, 010

\bibitem[\protect\citeauthoryear{Blazek, Vlah  \& Seljak}{Blazek
  et~al.}{2015}]{blazek_tidal_2015}
Blazek J.,  Vlah Z.,   Seljak U.,  2015, \mn@doi [\jcap]
  {10.1088/1475-7516/2015/08/015}, 8, 015

\bibitem[\protect\citeauthoryear{Blazek, MacCrann, Troxel  \& Fang}{Blazek
  et~al.}{2017}]{blazek_beyond_2017}
Blazek J.,  MacCrann N.,  Troxel M.~A.,   Fang X.,  2017, arXiv:1708.09247
  [astro-ph]

\bibitem[\protect\citeauthoryear{Chisari et~al.,}{Chisari
  et~al.}{2015}]{chisari_intrinsic_2015}
Chisari N.~E.,  et~al., 2015, arXiv:1507.07843 [astro-ph]

\bibitem[\protect\citeauthoryear{Clifton, Ferreira, Padilla  \&
  Skordis}{Clifton et~al.}{2012}]{clifton_modified_2012}
Clifton T.,  Ferreira P.~G.,  Padilla A.,   Skordis C.,  2012, \mn@doi
  [\physrep] {10.1016/j.physrep.2012.01.001}, 513, 1

\bibitem[\protect\citeauthoryear{Cole et~al.,}{Cole
  et~al.}{2005}]{cole_2df_2005}
Cole S.,  et~al., 2005, \mn@doi [\mnras] {10.1111/j.1365-2966.2005.09318.x},
  362, 505

\bibitem[\protect\citeauthoryear{Copeland, Sami  \& Tsujikawa}{Copeland
  et~al.}{2006}]{copeland_dynamics_2006}
Copeland E.~J.,  Sami M.,   Tsujikawa S.,  2006, \mn@doi [International Journal
  of Modern Physics D] {10.1142/S021827180600942X}, 15, 1753

\bibitem[\protect\citeauthoryear{Creminelli \& Vernizzi}{Creminelli \&
  Vernizzi}{2017}]{creminelli_dark_2017}
Creminelli P.,  Vernizzi F.,  2017, \mn@doi [Physical Review Letters]
  {10.1103/PhysRevLett.119.251302}, 119, 251302

\bibitem[\protect\citeauthoryear{Crittenden, Natarajan, Pen  \&
  Theuns}{Crittenden et~al.}{2001}]{crittenden_spin-induced_2001}
Crittenden R.~G.,  Natarajan P.,  Pen U.-L.,   Theuns T.,  2001, \mn@doi [\apj]
  {10.1086/322370}, 559, 552

\bibitem[\protect\citeauthoryear{{Daniel}, {Caldwell}, {Cooray}, {Serra}  \&
  {Melchiorri}}{{Daniel} et~al.}{2009}]{2009PhRvD..80b3532D}
{Daniel} S.~F.,  {Caldwell} R.~R.,  {Cooray} A.,  {Serra} P.,   {Melchiorri}
  A.,  2009, \mn@doi [\prd] {10.1103/PhysRevD.80.023532}, \href
  {http://esoads.eso.org/abs/2009PhRvD..80b3532D} {80, 023532}

\bibitem[\protect\citeauthoryear{Deffayet, Gao, Steer  \& Zahariade}{Deffayet
  et~al.}{2011}]{deffayet_$k$-essence_2011}
Deffayet C.,  Gao X.,  Steer D.~A.,   Zahariade G.,  2011, \mn@doi [Phys. Rev.
  D] {10.1103/PhysRevD.84.064039}, 84, 064039

\bibitem[\protect\citeauthoryear{Dossett, Ishak, Parkinson  \& Davis}{Dossett
  et~al.}{2015}]{dossett_constraints_2015}
Dossett J.~N.,  Ishak M.,  Parkinson D.,   Davis T.~M.,  2015, \mn@doi
  [Physical Review D] {10.1103/PhysRevD.92.023003}, 92

\bibitem[\protect\citeauthoryear{Ezquiaga \& Zumalac{\'a}rregui}{Ezquiaga \&
  Zumalac{\'a}rregui}{2017}]{ezquiaga_dark_2017}
Ezquiaga J.~M.,  Zumalac{\'a}rregui M.,  2017, \mn@doi [Physical Review
  Letters] {10.1103/PhysRevLett.119.251304}, 119, 251304

\bibitem[\protect\citeauthoryear{Falck, Koyama, Zhao  \& Li}{Falck
  et~al.}{2014}]{falck_vainshtein_2014}
Falck B.,  Koyama K.,  Zhao G.-b.,   Li B.,  2014, \mn@doi [JCAP]
  {10.1088/1475-7516/2014/07/058}, 2014, 058

\bibitem[\protect\citeauthoryear{Fortuna, Hoekstra, Joachimi, Johnston,
  Chisari, Georgiou  \& Mahony}{Fortuna et~al.}{2020}]{fortuna_halo_2020}
Fortuna M.~C.,  Hoekstra H.,  Joachimi B.,  Johnston H.,  Chisari N.~E.,
  Georgiou C.,   Mahony C.,  2020, arXiv:2003.02700 [astro-ph]

\bibitem[\protect\citeauthoryear{Fosalba \& Dor{\'e}}{Fosalba \&
  Dor{\'e}}{2007}]{fosalba_probing_2007}
Fosalba P.,  Dor{\'e} O.,  2007, \mn@doi [PRD] {10.1103/PhysRevD.76.103523},
  76, 103523

\bibitem[\protect\citeauthoryear{{Ghosh}, {Durrer}  \& {Schaefer}}{{Ghosh}
  et~al.}{2020}]{2020arXiv200504604G}
{Ghosh} B.,  {Durrer} R.,   {Schaefer} B.~M.,  2020, arXiv e-prints, \href
  {https://ui.adsabs.harvard.edu/abs/2020arXiv200504604G} {p. arXiv:2005.04604}

\bibitem[\protect\citeauthoryear{Giannantonio, Martinelli, Silvestri  \&
  Melchiorri}{Giannantonio et~al.}{2010}]{giannantonio_new_2010}
Giannantonio T.,  Martinelli M.,  Silvestri A.,   Melchiorri A.,  2010, Journal
  of Cosmology and Astroparticle Physics, 2010, 030

\bibitem[\protect\citeauthoryear{Gleyzes, Langlois  \& Vernizzi}{Gleyzes
  et~al.}{2014}]{gleyzes_unifying_2014}
Gleyzes J.,  Langlois D.,   Vernizzi F.,  2014, \mn@doi [International Journal
  of Modern Physics D] {10.1142/S021827181443010X}, 23, 1443010

\bibitem[\protect\citeauthoryear{Heavens, Matarrese  \& Verde}{Heavens
  et~al.}{1998}]{heavens_nonlinear_1998}
Heavens A.~F.,  Matarrese S.,   Verde L.,  1998, \mn@doi [MNRAS]
  {10.1046/j.1365-8711.1998.02052.x}, 301, 797

\bibitem[\protect\citeauthoryear{Heavens, Kitching  \& Verde}{Heavens
  et~al.}{2007}]{heavens_model_2007}
Heavens A.~F.,  Kitching T.~D.,   Verde L.,  2007, Mon. Not. Roy. Astron. Soc.,
  380, 1029

\bibitem[\protect\citeauthoryear{Heymans \& {others}}{Heymans \&
  {others}}{2013}]{heymans_cfhtlens_2013}
Heymans C.,  {others} 2013, \mn@doi [MNRAS] {10.1093/mnras/stt601}, 432, 2433

\bibitem[\protect\citeauthoryear{Hilbert, Xu, Schneider, Springel, Vogelsberger
   \& Hernquist}{Hilbert et~al.}{2016}]{hilbert_intrinsic_2016-1}
Hilbert S.,  Xu D.,  Schneider P.,  Springel V.,  Vogelsberger M.,   Hernquist
  L.,  2016, arXiv:1606.03216 [astro-ph], 468, 790

\bibitem[\protect\citeauthoryear{Hinshaw et~al.,}{Hinshaw
  et~al.}{2013}]{hinshaw_nine-year_2013}
Hinshaw G.,  et~al., 2013, \mn@doi [\apjs] {10.1088/0067-0049/208/2/19}, 208,
  19

\bibitem[\protect\citeauthoryear{Hirata \& Seljak}{Hirata \&
  Seljak}{2004}]{hirata_intrinsic_2004}
Hirata C.~M.,  Seljak U.,  2004, \mn@doi [Physical Review D]
  {10.1103/PhysRevD.70.063526}, 70, 063526

\bibitem[\protect\citeauthoryear{Hirata \& Seljak}{Hirata \&
  Seljak}{2010}]{hirata_intrinsic_2010}
Hirata C.~M.,  Seljak U.,  2010, \mn@doi [Physical Review D]
  {10.1103/PhysRevD.82.049901}, 82

\bibitem[\protect\citeauthoryear{Hirata, Mandelbaum, Ishak, Seljak, Nichol,
  Pimbblet, Ross  \& Wake}{Hirata et~al.}{2007}]{hirata_intrinsic_2007}
Hirata C.~M.,  Mandelbaum R.,  Ishak M.,  Seljak U.,  Nichol R.,  Pimbblet
  K.~A.,  Ross N.~P.,   Wake D.,  2007, \mn@doi [Monthly Notices of the Royal
  Astronomical Society] {10.1111/j.1365-2966.2007.12312.x}, 381, 1197

\bibitem[\protect\citeauthoryear{Hoekstra \& Jain}{Hoekstra \&
  Jain}{2008}]{hoekstra_weak_2008}
Hoekstra H.,  Jain B.,  2008, \mn@doi [Annual Review of Nuclear and Particle
  Science] {10.1146/annurev.nucl.58.110707.171151}, 58, 99

\bibitem[\protect\citeauthoryear{Horndeski}{Horndeski}{1974}]{horndeski_second-order_1974}
Horndeski G.~W.,  1974, \mn@doi [International Journal of Theoretical Physics]
  {10.1007/BF01807638}, 10, 363

\bibitem[\protect\citeauthoryear{Jain \& Zhang}{Jain \&
  Zhang}{2008}]{jain_observational_2008}
Jain B.,  Zhang P.,  2008, \mn@doi [\prd] {10.1103/PhysRevD.78.063503}, 78,
  063503

\bibitem[\protect\citeauthoryear{Joachimi \& Schneider}{Joachimi \&
  Schneider}{2010}]{joachimi_intrinsic_2010}
Joachimi B.,  Schneider P.,  2010, \mn@doi [\aap]
  {10.1051/0004-6361/201014482}, 517, A4

\bibitem[\protect\citeauthoryear{Joachimi, Mandelbaum, Abdalla  \&
  Bridle}{Joachimi et~al.}{2011}]{joachimi_constraints_2011}
Joachimi B.,  Mandelbaum R.,  Abdalla F.~B.,   Bridle S.~L.,  2011, \mn@doi
  [\aap] {10.1051/0004-6361/201015621}, 527, A26

\bibitem[\protect\citeauthoryear{Joachimi, Semboloni, Bett, Hartlap, Hilbert,
  Hoekstra, Schneider  \& Schrabback}{Joachimi
  et~al.}{2013a}]{joachimi_intrinsic_2013-1}
Joachimi B.,  Semboloni E.,  Bett P.~E.,  Hartlap J.,  Hilbert S.,  Hoekstra
  H.,  Schneider P.,   Schrabback T.,  2013a, \mn@doi [\mnras]
  {10.1093/mnras/stt172}, 431, 477

\bibitem[\protect\citeauthoryear{Joachimi, Semboloni, Hilbert, Bett, Hartlap,
  Hoekstra  \& Schneider}{Joachimi et~al.}{2013b}]{joachimi_intrinsic_2013}
Joachimi B.,  Semboloni E.,  Hilbert S.,  Bett P.~E.,  Hartlap J.,  Hoekstra
  H.,   Schneider P.,  2013b, \mn@doi [\mnras] {10.1093/mnras/stt1618}, 436,
  819

\bibitem[\protect\citeauthoryear{Johnston et~al.,}{Johnston
  et~al.}{2019}]{johnston_kidsgama_2019}
Johnston H.,  et~al., 2019, \mn@doi [Astronomy and Astrophysics]
  {10.1051/0004-6361/201834714}, 624, A30

\bibitem[\protect\citeauthoryear{Joyce, Lombriser  \& Schmidt}{Joyce
  et~al.}{2016}]{joyce_dark_2016}
Joyce A.,  Lombriser L.,   Schmidt F.,  2016, \mn@doi [Annual Review of Nuclear
  and Particle Science] {10.1146/annurev-nucl-102115-044553}, 66, 95

\bibitem[\protect\citeauthoryear{Kiessling et~al.,}{Kiessling
  et~al.}{2015}]{kiessling_galaxy_2015-1}
Kiessling A.,  et~al., 2015, preprint

\bibitem[\protect\citeauthoryear{Kilbinger}{Kilbinger}{2015}]{kilbinger_cosmology_2015}
Kilbinger M.,  2015, \mn@doi [Reports on Progress in Physics]
  {10.1088/0034-4885/78/8/086901}, 78, 086901

\bibitem[\protect\citeauthoryear{Kim \& Linder}{Kim \&
  Linder}{2019}]{kim_complementarity_2019}
Kim A.~G.,  Linder E.~V.,  2019, preprint (\mn@eprint {} {1911.09121})

\bibitem[\protect\citeauthoryear{Kirk et~al.,}{Kirk
  et~al.}{2015}]{kirk_galaxy_2015}
Kirk D.,  et~al., 2015, \mn@doi [Space Science Reviews]
  {10.1007/s11214-015-0213-4}, 193, 139

\bibitem[\protect\citeauthoryear{Kobayashi, Tashiro  \& Suzuki}{Kobayashi
  et~al.}{2010}]{kobayashi_evolution_2010}
Kobayashi T.,  Tashiro H.,   Suzuki D.,  2010, \mn@doi [Physical Review D]
  {10.1103/PhysRevD.81.063513}, 81

\bibitem[\protect\citeauthoryear{Koyama}{Koyama}{2016}]{koyama_cosmological_2016}
Koyama K.,  2016, \mn@doi [Reports on Progress in Physics]
  {10.1088/0034-4885/79/4/046902}, 79, 046902

\bibitem[\protect\citeauthoryear{Kraljic, Dav{\'e}  \& Pichon}{Kraljic
  et~al.}{2020}]{kraljic_and_2020}
Kraljic K.,  Dav{\'e} R.,   Pichon C.,  2020, \mn@doi [Monthly Notices of the
  Royal Astronomical Society] {10.1093/mnras/staa250}, 493, 362

\bibitem[\protect\citeauthoryear{Kunz \& Sapone}{Kunz \&
  Sapone}{2007}]{kunz_dark_2007}
Kunz M.,  Sapone D.,  2007, \mn@doi [Phys.Rev.Lett.]
  {10.1103/PhysRevLett.98.121301}, 98, 121301

\bibitem[\protect\citeauthoryear{{LSST Dark Energy Science
  Collaboration}}{{LSST Dark Energy Science
  Collaboration}}{2012}]{lsst_dark_energy_science_collaboration_large_2012}
{LSST Dark Energy Science Collaboration} 2012, preprint

\bibitem[\protect\citeauthoryear{Laszlo \& Bean}{Laszlo \&
  Bean}{2007}]{laszlo_non-linear_2007}
Laszlo I.,  Bean R.,  2007, ArXiv e-prints 0709.0307, 709

\bibitem[\protect\citeauthoryear{Laureijs et~al.,}{Laureijs
  et~al.}{2011}]{laureijs_euclid_2011}
Laureijs R.,  et~al., 2011, preprint

\bibitem[\protect\citeauthoryear{Lesgourgues}{Lesgourgues}{2011}]{lesgourgues_cosmic_2011}
Lesgourgues J.,  2011, preprint

\bibitem[\protect\citeauthoryear{Limber}{Limber}{1954}]{limber_analysis_1954}
Limber D.~N.,  1954, \apj, 119, 655

\bibitem[\protect\citeauthoryear{Linder, Seng{\"o}r  \& Watson}{Linder
  et~al.}{2016}]{linder_is_2016}
Linder E.~V.,  Seng{\"o}r G.,   Watson S.,  2016, \mn@doi [\jcap]
  {10.1088/1475-7516/2016/05/053}, 5, 053

\bibitem[\protect\citeauthoryear{Lombriser \& Lima}{Lombriser \&
  Lima}{2017}]{lombriser_challenges_2017}
Lombriser L.,  Lima N.~A.,  2017, \mn@doi [Physics Letters B]
  {10.1016/j.physletb.2016.12.048}, 765, 382

\bibitem[\protect\citeauthoryear{Loverde \& Afshordi}{Loverde \&
  Afshordi}{2008}]{loverde_extended_2008}
Loverde M.,  Afshordi N.,  2008, \mn@doi [\prd] {10.1103/PhysRevD.78.123506},
  78, 123506

\bibitem[\protect\citeauthoryear{Lue, Scoccimarro  \& Starkman}{Lue
  et~al.}{2004}]{lue_differentiating_2004}
Lue A.,  Scoccimarro R.,   Starkman G.,  2004, \mn@doi [\prd]
  {10.1103/PhysRevD.69.044005}, 69, 044005

\bibitem[\protect\citeauthoryear{Majerotto et~al.,}{Majerotto
  et~al.}{2012}]{majerotto_probing_2012}
Majerotto E.,  et~al., 2012, MNRAS, 424, 1392

\bibitem[\protect\citeauthoryear{Mandelbaum, Hirata, Ishak, Seljak  \&
  Brinkmann}{Mandelbaum et~al.}{2006}]{mandelbaum_detection_2006}
Mandelbaum R.,  Hirata C.~M.,  Ishak M.,  Seljak U.,   Brinkmann J.,  2006,
  \mn@doi [\mnras] {10.1111/j.1365-2966.2005.09946.x}, 367, 611

\bibitem[\protect\citeauthoryear{Mortonson, Weinberg  \& White}{Mortonson
  et~al.}{2013}]{mortonson_dark_2013}
Mortonson M.~J.,  Weinberg D.~H.,   White M.,  2013, arXiv:1401.0046 [astro-ph]

\bibitem[\protect\citeauthoryear{Nicolis, Rattazzi  \& Trincherini}{Nicolis
  et~al.}{2009}]{nicolis_galileon_2009}
Nicolis A.,  Rattazzi R.,   Trincherini E.,  2009, \mn@doi [\prd]
  {10.1103/PhysRevD.79.064036}, 79, 064036

\bibitem[\protect\citeauthoryear{Okumura \& Jing}{Okumura \&
  Jing}{2009}]{okumura_gravitational_2009}
Okumura T.,  Jing Y.~P.,  2009, \mn@doi [\apjl] {10.1088/0004-637X/694/1/L83},
  694, L83

\bibitem[\protect\citeauthoryear{Perlmutter et~al.,}{Perlmutter
  et~al.}{1998}]{perlmutter_discovery_1998}
Perlmutter S.,  et~al., 1998, \mn@doi [\nat] {10.1038/34124}, 391, 51

\bibitem[\protect\citeauthoryear{Perlmutter, Aldering, Goldhaber  \& {et
  al.}}{Perlmutter et~al.}{1999}]{perlmutter_measurements_1999}
Perlmutter S.,  Aldering G.,  Goldhaber G.,   {et al.} 1999, \mn@doi [\apj]
  {10.1086/307221}, 517, 565

\bibitem[\protect\citeauthoryear{{Planck Collaboration} et~al.,}{{Planck
  Collaboration} et~al.}{2018}]{planck_collaboration_planck_2018}
{Planck Collaboration} et~al., 2018, arXiv e-prints

\bibitem[\protect\citeauthoryear{{Planck Collaboration} et~al.,}{{Planck
  Collaboration} et~al.}{2020}]{planck_collaboration_planck_2020}
{Planck Collaboration} et~al., 2020, \mn@doi [Astronomy and Astrophysics]
  {10.1051/0004-6361/201833910}, 641, A6

\bibitem[\protect\citeauthoryear{Reischke \& Sch{\"a}fer}{Reischke \&
  Sch{\"a}fer}{2019}]{reischke_environmental_2019}
Reischke R.,  Sch{\"a}fer B.~M.,  2019, \mn@doi [Journal of Cosmology and
  Astroparticle Physics] {10.1088/1475-7516/2019/04/031}, 04, 031

\bibitem[\protect\citeauthoryear{Reischke, Mancini, Sch{\"a}fer  \&
  Merkel}{Reischke et~al.}{2019}]{reischke_investigating_2019}
Reischke R.,  Mancini A.~S.,  Sch{\"a}fer B.~M.,   Merkel P.~M.,  2019, \mn@doi
  [Monthly Notices of the Royal Astronomical Society] {10.1093/mnras/sty2919},
  482, 3274

\bibitem[\protect\citeauthoryear{Riess, Filippenko, Challis  \& {et al.}}{Riess
  et~al.}{1998}]{riess_observational_1998}
Riess A.~G.,  Filippenko A.~V.,  Challis P.,   {et al.} 1998, \mn@doi [\aj]
  {10.1086/300499}, 116, 1009

\bibitem[\protect\citeauthoryear{Riess, Strolger, Tonry, Casertano, Ferguson
  \& {et al.}}{Riess et~al.}{2004}]{riess_type_2004}
Riess A.~G.,  Strolger L.-G.,  Tonry J.,  Casertano S.,  Ferguson H.~C.,   {et
  al.} 2004, \mn@doi [\apj] {10.1086/383612}, 607, 665

\bibitem[\protect\citeauthoryear{Riess, Strolger, Casertano, Ferguson,
  Mobasher, Gold, Challis  \& {et al.}}{Riess et~al.}{2007}]{riess_new_2007}
Riess A.~G.,  Strolger L.-G.,  Casertano S.,  Ferguson H.~C.,  Mobasher B.,
  Gold B.,  Challis P.~J.,   {et al.} 2007, \mn@doi [\apj] {10.1086/510378},
  659, 98

\bibitem[\protect\citeauthoryear{Riess, Casertano, Yuan, Macri  \&
  Scolnic}{Riess et~al.}{2019}]{riess_large_2019}
Riess A.~G.,  Casertano S.,  Yuan W.,  Macri L.~M.,   Scolnic D.,  2019,
  arXiv:1903.07603 [astro-ph]

\bibitem[\protect\citeauthoryear{Sakstein \& Jain}{Sakstein \&
  Jain}{2017}]{sakstein_implications_2017}
Sakstein J.,  Jain B.,  2017, \mn@doi [Physical Review Letters]
  {10.1103/PhysRevLett.119.251303}, 119, 251303

\bibitem[\protect\citeauthoryear{Samuroff, Mandelbaum  \& Blazek}{Samuroff
  et~al.}{2020}]{samuroff_advances_2020}
Samuroff S.,  Mandelbaum R.,   Blazek J.,  2020, arXiv e-prints, 2009,
  arXiv:2009.10735

\bibitem[\protect\citeauthoryear{Satpathy et~al.,}{Satpathy
  et~al.}{2017}]{satpathy_clustering_2017}
Satpathy S.,  et~al., 2017, \mn@doi [Monthly Notices of the Royal Astronomical
  Society] {10.1093/mnras/stx883}, 469, 1369

\bibitem[\protect\citeauthoryear{Schaefer}{Schaefer}{2009}]{schaefer_review:_2009}
Schaefer B.~M.,  2009, \mn@doi [International Journal of Modern Physics D]
  {10.1142/S0218271809014388}, 18, 173

\bibitem[\protect\citeauthoryear{Shapiro, Crittenden  \& Percival}{Shapiro
  et~al.}{2012}]{shapiro_complementarity_2012}
Shapiro C.,  Crittenden R.~G.,   Percival W.~J.,  2012, \mn@doi [MNRAS]
  {10.1111/j.1365-2966.2012.20785.x}, 422, 2341

\bibitem[\protect\citeauthoryear{Sipp, Schaefer  \& Reischke}{Sipp
  et~al.}{2020}]{sipp_optimising_2020}
Sipp M.,  Schaefer B.~M.,   Reischke R.,  2020, arXiv e-prints, 2002,
  arXiv:2002.12695

\bibitem[\protect\citeauthoryear{Spurio~Mancini, Reischke, Pettorino,
  Sch{\"a}fer  \& Zumalac{\'a}rregui}{Spurio~Mancini
  et~al.}{2018}]{spurio_mancini_testing_2018}
Spurio~Mancini A.,  Reischke R.,  Pettorino V.,  Sch{\"a}fer B.~M.,
  Zumalac{\'a}rregui M.,  2018, \mn@doi [Monthly Notices of the Royal
  Astronomical Society] {10.1093/mnras/sty2092}, 480, 3725

\bibitem[\protect\citeauthoryear{Tegmark, Taylor  \& Heavens}{Tegmark
  et~al.}{1997}]{tegmark_karhunen-loeve_1997-1}
Tegmark M.,  Taylor A.,   Heavens A.,  1997, \mn@doi [Astrophys. J.]
  {10.1086/303939}, 480, 22

\bibitem[\protect\citeauthoryear{Tenneti, Singh, Mandelbaum, Matteo, Feng  \&
  Khandai}{Tenneti et~al.}{2015a}]{tenneti_intrinsic_2015}
Tenneti A.,  Singh S.,  Mandelbaum R.,  Matteo T.~D.,  Feng Y.,   Khandai N.,
  2015a, \mn@doi [\mnras] {10.1093/mnras/stv272}, 448, 3522

\bibitem[\protect\citeauthoryear{Tenneti, Mandelbaum, Di~Matteo, Kiessling  \&
  Khandai}{Tenneti et~al.}{2015b}]{tenneti_galaxy_2015}
Tenneti A.,  Mandelbaum R.,  Di~Matteo T.,  Kiessling A.,   Khandai N.,  2015b,
  \mn@doi [Monthly Notices of the Royal Astronomical Society]
  {10.1093/mnras/stv1625}, 453, 469

\bibitem[\protect\citeauthoryear{Troxel \& Ishak}{Troxel \&
  Ishak}{2015}]{troxel_intrinsic_2015}
Troxel M.~A.,  Ishak M.,  2015, \mn@doi [Physics Reports]
  {10.1016/j.physrep.2014.11.001}, 558, 1

\bibitem[\protect\citeauthoryear{Tugendhat \& Schaefer}{Tugendhat \&
  Schaefer}{2017}]{tugendhat_angular_2017}
Tugendhat T.~M.,  Schaefer B.~M.,  2017, arXiV

\bibitem[\protect\citeauthoryear{Vlah, Chisari  \& Schmidt}{Vlah
  et~al.}{2019}]{vlah_eft_2019}
Vlah Z.,  Chisari N.~E.,   Schmidt F.,  2019, arXiv e-prints, 1910,
  arXiv:1910.08085

\bibitem[\protect\citeauthoryear{White}{White}{2016}]{white_marked_2016}
White M.,  2016, arXiv:1609.08632 [astro-ph]

\bibitem[\protect\citeauthoryear{Zhao, Pogosian, Silvestri  \& Zylberberg}{Zhao
  et~al.}{2009}]{zhao_searching_2009}
Zhao G.-B.,  Pogosian L.,  Silvestri A.,   Zylberberg J.,  2009, \mn@doi
  [Physical Review D] {10.1103/PhysRevD.79.083513}, 79

\bibitem[\protect\citeauthoryear{Zjupa, Sch{\"a}fer  \& Hahn}{Zjupa
  et~al.}{2020}]{zjupa_intrinsic_2020}
Zjupa J.,  Sch{\"a}fer B.~M.,   Hahn O.,  2020, arXiv:2010.07951 [astro-ph]

\bibitem[\protect\citeauthoryear{Zumalacarregui, Bellini, Sawicki  \&
  Lesgourgues}{Zumalacarregui et~al.}{2016}]{zumalacarregui_hiclass:_2016}
Zumalacarregui M.,  Bellini E.,  Sawicki I.,   Lesgourgues J.,  2016, preprint

\makeatother
\end{thebibliography}

\bsp
\label{lastpage}
\end{document}